\documentclass[review]{elsarticle}

\usepackage{lineno,hyperref}
\modulolinenumbers[5]

\usepackage{upgreek}
\usepackage{subfigure}
\usepackage{textcomp}
\usepackage{amssymb}
\usepackage{amsthm}

\begin{document}

\begin{frontmatter}

\title{A comparative study of alternative methods to determine the response of poly-ethylene terephthalate nuclear track detector}

\author[mymainaddress]{R. Bhattacharyya\corref{mycorrespondingauthor}}
\cortext[mycorrespondingauthor]{Corresponding author}
\ead{rupamoy@gmail.com}

\author[mymainaddress]{S. Dey}

\author[mymainaddress,mysecondaryaddress]{Sanjay K. Ghosh}
\author[mytertiaryaddress]{Akhil Jhingan}

\author[mymainaddress]{A. Maulik}

\author[myquaternaryaddress]{L. Patrizii}

\author[mymainaddress,mysecondaryaddress]{Sibaji Raha}

\author[mymainaddress]{D. Syam}
\author[myquaternaryaddress]{V. Togo}

\address[mymainaddress]{Centre for Astroparticle Physics and Space Science, Bose Institute, Kolkata 700 091, India}
\address[mysecondaryaddress]{Department of Physics, Bose Institute, Kolkata 700 009, India}
\address[mytertiaryaddress]{Inter University Accelerator Centre, Aruna Asaf Ali Marg, New Delhi 110067, India}
\address[myquaternaryaddress]{INFN, Sezione di Bologna, Viale C. Berti Pichat 6/2, I-40127 Bologna, Italy}

\begin{abstract}
Two widely used methods of determining the etch-rate ratio in poly-ethylene terephthalate (PET)
nuclear track detector are compared. Their application in 
different regimes of ion's energy loss is investigated. A new calibration curve for PET
is also presented. 
\end{abstract}

\begin{keyword}
Nuclear Track Detector\sep Poly-ethylene Terephthalate\sep Restricted Energy Loss
\end{keyword}

\end{frontmatter}


\section{Introduction}

Nuclear track detectors (NTDs) find wide use in charged particle
detection~\cite{Fleischer:1975ya,Durrani:1987zv}. They are particularly suited in the search for rare, highly ionizing
particles against a large background from low ionizing 
particles~\cite{Balestra:2008ps,Cecchini:2008su,Cecchini:2009br,Acharya:2014nyr,PhysRevLett.85.1384,1475-7516-2017-04-035}. 

A charged particle losing
energy while passing through an NTD, can create a permanent
trail (\textquotedblleft latent track\textquotedblright) of damaged bonds along its path~\cite{Fleischer:1975ya}. On being treated with suitable
reagents (chemical etching), if the material along the damage trail is etched out at a faster rate (the track etch-rate, $V_T$)
than the etch-rate of the undamaged
material (the bulk etch-rate, $V_B$), a conical etch-pit is formed.
In nuclear track detectors the size and shape of etch-pits depend on the particle's energy loss - more precisely 
on the restricted energy loss ($REL$) - along its path and on its angle of incidence~\cite{BENTON1969343,BAIOCCHI1995145}.
The minimum $REL$ allowing the formation of an etch-pit, i.e. such that $V_T>V_B$, sets the detector threshold.
\begin{figure} 
 \centering
  \subfigure[]{
    \includegraphics[width=271.6px,height=232.75px]{normal_Vt_const.eps}   
  } 
  \quad 
  \subfigure[]{ 
    \includegraphics[width=271.6px,height=232.75px]{normal_Vt_change.eps}  
  }
    \caption{Sketch of an etch-pit along the latent track of an ion incident normally to the NTD's surface. The energy loss is 
    (a) constant, (b) increasing along the ion's trajectory.
Here $\Delta t$ is the etching time at which the etchant reaches the level of the post-etch surface along the track.} 
    \label{normal}
\end{figure}
For a particle impinging perpendicularly to the detector surface, if its energy loss is above threshold and constant, 
the etch-pit has the shape of a right circular cone. The process is
similar to the Mach cone produced by an object moving in a medium at constant supersonic velocity.
$V_T$ and $V_B$ play the role of the velocity of the object and that of the sound in the medium, respectively.
We can identify the angle $\alpha$ (Fig.~\ref{normal}a) with the Mach angle and $\sin\alpha=V_B/V_T$.
The shape of the etch-pit will not be a right-circular cone (or obliquely cut right-circular cone in case of angular incidence) 
if $dE/dx$ (as well as $REL$) varies along the latent track.
The steepness of the etch-pit's wall increases (Fig.~\ref{normal}b) until the energy loss reaches the Bragg peak~\cite{Fleischer:1975ya}. Thus 
the etch-pit will look like a flared 
cone~\cite{NIKEZIC200451} as sketched in Fig.~\ref{normal}b. 
\par 

The aim of this paper is to determine 
the etch-rate ratio of poly-ethylene terephthalate (PET) for ions with energy in the range $0.7-11.1$ MeV/nucleon.

\section{Methods to determine the etch-rate ratio}
Different methods can be used to calculate the etch-rate ratio $V=V_T/V_B$ from the dimensions of the etch-pits. In this paper, 
two such methods are compared, which, in general, relate to different regimes of the ion's energy loss. 
By one method~\cite{BHOWMIK2011197} 
$V_T$ is calculated by measuring the length $L_h$ (Fig.~\ref{normal}a) of 
the etched out section of the latent track and the etching time ($t$). The bulk etch-rate $V_B$ is determined by the change 
in the thickness of the detector sheet
over the
etching time. With reference to Fig.~\ref{oblique}a, for a particle impinging at an angle $i$ with respect to the normal to the detector 
surface, one has
\begin{equation}
\label{lab}
\frac{V_T}{V_B}=\frac{d+V_Bt}{V_Bt~\cos i}~~(i<\alpha)~;~~\frac{V_T}{V_B}=\frac{\mu d'+V_Bt}{V_Bt~\cos i}~~(i>\alpha)
\end{equation}
where $d$ (or $d'$) is the vertical distance between the post-etching surface and the tip of the etch-pit cone, as measured by the microscope.
The refractive index of the detector 
material, $\mu$, has to be taken into account for $i>\alpha$ (Fig.~\ref{oblique}b). 
The refractive index of PET is $\mu_{PET}=1.64\pm0.02$ in yellow light. It was determined 
following the same procedure as in~\cite{BALESTRA2007254} 
and it is identical to 
the value given in~\cite{ELMAN1998814}. Henceforth this method of determining $V_T/V_B$ is referred to as \textquotedblleft depth
measurement method\textquotedblright.
\begin{figure} 
 \centering
  \subfigure[]{%
    \includegraphics[width=252.2px,height=108.3px]{d_oblique.eps}  
  } 
  \quad 
  \subfigure[]{%
    \includegraphics[width=252.2px,height=108.3px]{nd_oblique.eps}  
  }
    \caption{Etched latent track when the ion impinges on the NTD's surface at an angle (a) $i<\alpha$ and (b) $i>\alpha$.} 
    \label{oblique}
\end{figure}
\par
By another method, the average $V_T/V_B$ can be determined from the size of the surface etch-pit opening~\cite{SOMOGYI1973211,Pinfold:2009oia}, using Eq.~(\ref{MoEDAL})
\begin{equation}
\label{MoEDAL}
\frac{V_T}{V_B}=\sqrt{1+\frac{4A^2}{(1-B^2)^2}}
\end{equation}
where $A=\frac{a}{V_Bt}$ and $B=\frac{b}{V_Bt}$; $a$ and $b$ are the semi-major and the semi-minor axis of the 
elliptical opening of the etch-pit, respectively. 
For a homogeneous and isotropic material, $B\leqslant1$. For a particle impinging normally to the detector surface, 
 the opening of the etch-pit is circular, and $a=b$. This method is referred
to in the following as \textquotedblleft diameter measurement method\textquotedblright.

\par

It has to be noted that Eq.~(\ref{MoEDAL}) provides the average of $V_T/V_B$ over the length $L_d=V_BtV/(V+1)$~\cite{BAIOCCHI1995145} 
from the pre-etching 
surface (Fig.~\ref{normal}a), whereas with Eq.~(\ref{lab}) the average is over the length $L_h=(d+V_Bt)/\cos i$ or $L_h=(\mu d'+V_Bt)/\cos i$.
Therefore for a slowing down electrically charged particle, the etch-rate ratio computed  using Eq.~(\ref{lab}) yields a value
larger than the one computed using Eq.~(\ref{MoEDAL}),
since $V_T/V_B$ is an increasing function of $REL$~\cite{Fleischer:1975ya,BALESTRA2007254,DEY2011805}.  

\section{Determination of \boldmath $REL$}

At the energies of ions used in this paper, the restricted energy loss for NTDs can be computed using Eq.~(\ref{REL})~\cite{Fleischer:1975ya}
\begin{equation}
\label{REL}
  \bigg(\frac{dE}{dx}\bigg)_{E<E_{cut}}=C_1\bigg(\frac{z^\ast}{\beta}\bigg)^2\bigg[\ln\bigg(\frac{W_{max}E_{cut}}{I^2}\bigg)-\beta^2-\delta\bigg]
\end{equation}
where $C_1=2\pi n_ee^4/mc^2$;
$n_e$ is the number density of electrons in the detector;
$m_e$ is the electron mass;
$z^\ast$ is the effective charge of the incoming particle~\cite{Fleischer:1975ya}
\begin{equation}
\label{charge}
z^\ast=z[1-e^{(-130\beta/z^{2/3})}]
\end{equation}
$z$ and $\beta$ being the particle's electric charge and velocity in units of electron charge $e$ and speed of light $c$, respectively; 
  $W_{max}=2m_ec^2\beta^2\gamma^2$ 
in the \textquotedblleft low-energy\textquotedblright~approximation~\cite{Rossi:1952ab}
is the maximum energy that can be transferred 
to an electron in a single collision; $\gamma$ is the Lorentz factor; 
$\delta$ is the density-effect correction term due to the polarization of the medium, relevant at relativistic energies;
$I$ is the material mean ionization potential ($I=73.2$ eV for PET~\cite{Michael:2013ab});
$E_{cut}$ is the maximum energy of delta-rays contributing to the formation of the latent track. 
Hereinafter, $E_{cut}= 350$ eV is assumed to compute $REL$ in poly-ethylene terephthalate. For calculating the average of 
$REL$ along the ion's trajectory, the Monte Carlo code
SRIM~\cite{Ziegler20101818} was used.

\section{Experimental method}
\begin{figure}[h]
\centering
\includegraphics[width=250px,height=200px]{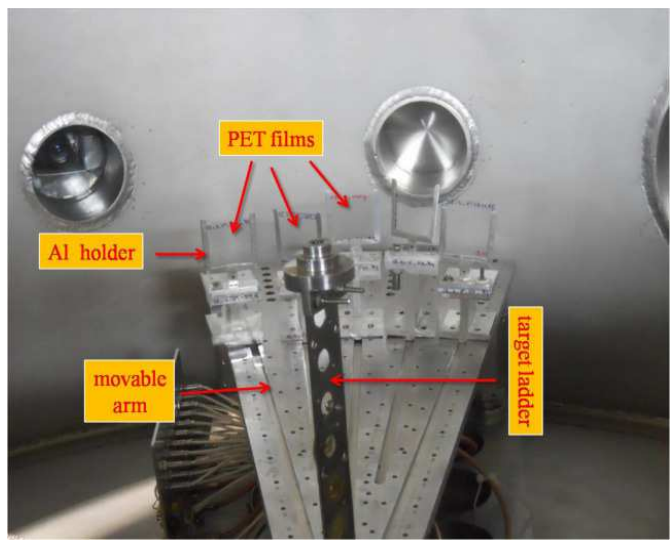}
\caption{Interiors of the General Purpose Scattering Chamber (diameter 1 m).}
\label{beam_IUAC} 
\end{figure}

In order to compare the \textquotedblleft depth
measurement method\textquotedblright~to the \textquotedblleft diameter measurement method\textquotedblright~when $B\approx1$,
we used 3.8 MeV/A $^{35}$Cl$^{10+}$ beam from the pelletron accelerator 
and the General Purpose Scattering Chamber at the Inter-University Accelerator Center (IUAC), New Delhi.
Details of the beam are given in Table~\ref{beam details_IUAC}. 
Small pieces (5 cm $\times$ 5 cm) of 90 $\upmu$m thick PET films (Desmat Co., India), were mounted on the aluminum 
 holders and placed on the two arms 
inside the scattering chamber (Fig.~\ref{beam_IUAC}). 
PET films were irradiated by 
$^{35}$Cl$^{10+}$ ions backscattered from a gold
 foil target 250 $\upmu$g cm$^{-2}$ thick.  
Exposure duration was controlled such that the ion 
 density on the detector never exceeded $\sim10^4$/cm$^2$
to prevent detector \textquotedblleft burnout\textquotedblright. After the exposure, the detectors were etched
in a tank (Julabo, Germany)
equipped with a motorized stirrer, in 6.25 N  NaOH aqueous solution at 55.0 $\pm$ 0.1$^{\circ}$C for 3 hours. After etching,
the detectors were observed under a Leica DM4000 B optical microscope with a 100x objective and 10x eyepieces. 
The image analysis software QWin was used for the measurement of etch-pits' sizes. 
\begin{table}
\centering
\begin{tabular}{|c|c|c|c|} \hline
\bf Ion       & \bf Energy per    & \bf Beam current  & \bf Charge state\\
              & \bf nucleus (MeV)    & \bf (nA)          &                 \\ \hline           
$~^{35}$Cl        & $132$          & $15$             & $10$           \\ \hline
\end{tabular}
\caption{Properties of the $^{35}$Cl$^{10+}$ beam. Beam energy has an uncertainty $<5\%$.}
\label{beam details_IUAC}
\end{table}
\section{Results}
\begin{figure}[h]
\centering
\includegraphics[width=350px,height=300px]{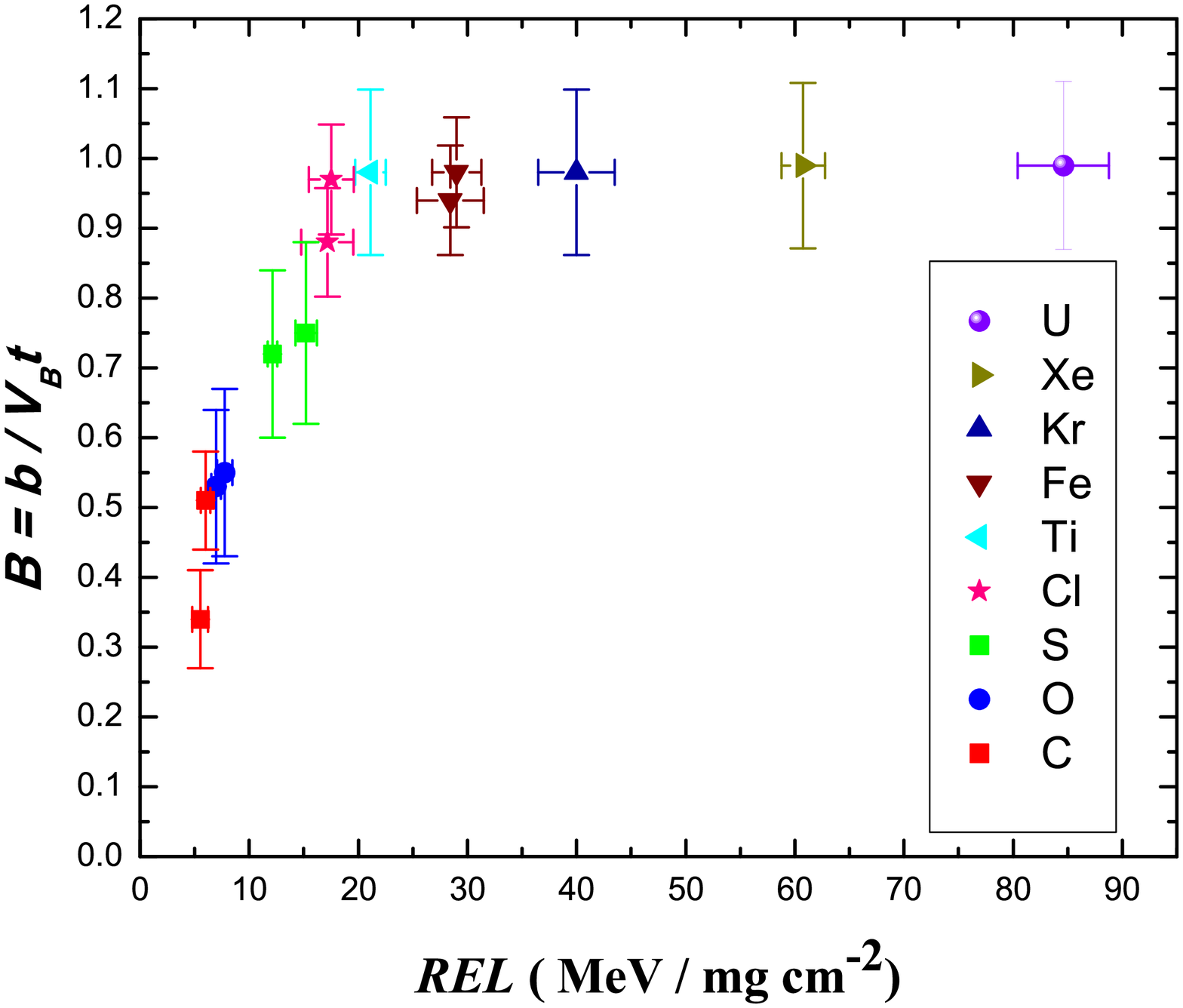}
\caption{Normalized semi minor-axis $B$ vs. $REL$ plot for different ions, listed in table 3.}
\label{Bcut} 
\end{figure}

Data obtained from previous exposures of PET films and from the exposure to $^{35}$Cl$^{10+}$ ions are summarized in 
Table~\ref{Data_Pitfall} and Table~\ref{ions_PET}. As shown in the last two rows of Table~\ref{ions_PET}  
the values of $V_T/V_B$ for $^{35}$Cl$^{10+}$ ions of 70 MeV$/$nucleus and 77 MeV$/$nucleus, determined
using Eq.~(\ref{MoEDAL}) are significantly different, although the corresponding values of $REL$ are very close. This happens because of 
the $(1-B^2)$ term in the denominator of Eq.~(\ref{MoEDAL}), when $B\sim1$ (columns 6 of table 3 and  Fig.~\ref{Bcut}). 
\begin{tiny} 
\begin{table}[!h]
\centering
\resizebox{\textwidth}{!}{
\begin{tabular}{|c|c|c|c|c|c|c|} \hline  
\bf Ion       & \bf  Energy  per    & \bf Incidence  & \bf Time of         &\bf Depth  & \bf Major-axis & \bf Minor-axis\\
              & \bf   nucleon (MeV/A)    & \bf angle      & \bf etching (hours) & \bf measurement \boldmath ($\upmu$m)& \bf \boldmath ($\upmu$m)& \bf \boldmath ($\upmu$m)  \\ \hline 

$~^{32}$S$^{9+}$&2.1		&0\textdegree    &1.8			&$6.53\pm0.30$		&$2.87\pm0.15$		&$2.75\pm0.14$\\
		&3.4		&0\textdegree&2.0			&$5.32\pm0.29$		&$2.91\pm0.14$		&$2.86\pm0.14$\\  \hline

$~^{16}$O$^{7+}$	&1.0		&0\textdegree&3.0		&$5.11\pm0.29$		&$3.62\pm0.15$		&$3.34\pm0.15$\\
		&1.2		&0\textdegree&3.0       		&$4.05\pm0.30$		& $3.54\pm0.16$		&$3.21\pm0.0.16$\\ \hline

$~^{12}$C$^{4+}$&0.7		&25\textdegree&4.0			&$5.12\pm0.29$		&$4.50\pm0.14$		&$4.09\pm0.14$\\
		&0.9		&25\textdegree&4.0			&$4.03\pm0.29$		&$3.46\pm0.14$		&$2.76\pm0.15$\\	\hline	\hline
	
$~^{35}$Cl$^{10+}$&2.0		&0\textdegree&3.0			&$12.25\pm0.29$		& $5.94\pm0.15$		&$5.87\pm0.15$\\
		&2.2		&0\textdegree&3.0			&$10.81\pm0.29$		& $5.29\pm0.14$		& $5.26\pm0.14$\\ \hline
\end{tabular}}
\caption{Data used to compute $V_T/V_B$ using Eq.~(\ref{lab}) and Eq.~(\ref{MoEDAL}).
Ions' energies and angles of incidences have an uncertainty $<10\%$ and $<4\%$, respectively.
Errors in column 5, 6 and 7 include statistical and systematic uncertainties.}
\label{Data_Pitfall}
\end{table}
\end{tiny}

\begin{tiny} 
\begin{table}[!h]
\centering
\resizebox{\textwidth}{!}{
\begin{tabular}{|c|c|c|c|c|c|c|c|} \hline
              & \bf Incident      & \bf Energy at   & \bf \boldmath Average $REL$                     & \bf \boldmath Average $REL$                     & \bf Normalized     & \boldmath $V_T/V_B$      & \boldmath $V_T/V_B$  \\ 
\bf Ion	      & \bf  energy per   & \bf the Bragg   & \bf over the                        & \bf over the                        & \bf  semi    & \bf by depth   & \bf by diameter \\ 
              & \bf nucleus       & \bf peak        & \bf length \boldmath $L_d$                    & \bf \boldmath length $L_h$ &  \bf minor-axis              & \bf measurement& \bf measurement \\ 
              &  \bf (MeV)        &    \bf (MeV)    & \bf \boldmath (MeV/mg cm${^{\textbf -2}}$)   & \bf \boldmath (MeV/mg cm${^{\textbf -2}}$)   &  \textbf{\textit{B}}& \bf method     & \bf method \\ \hline
 
$~^{32}$S$^{9+}$   &67           &      22.5        &$14.63\pm0.22$                       & $15.2\pm1.0$                          &$0.75\pm0.13$      & $4.5\pm0.7$  & $3.8\pm1.3$\\
             &110                &                &$11.86\pm0.13$                         &$12.1\pm0.4$                       &$0.72\pm0.12$        & $3.7\pm0.6$ &$3.2\pm0.9$\\ \hline
$~^{16}$O$^{7+}$    &16                & 7.0       &$7.34\pm0.25$                          & $7.8\pm0.7$                & $0.55\pm0.12$        & $2.7\pm0.4$ & $2.0\pm0.5$\\
             &20                 &                &$6.67\pm0.15$                           & $7.0\pm0.4$                   &$0.53\pm0.11$          & $2.3\pm0.4$& $1.9\pm0.4$\\  \hline 
$~^{12}$C$^{4+}$    &8                  & 5.0    & $5.9\pm0.5$                          &$6.0\pm0.4$                & $0.51\pm0.07$         & $2.1\pm0.5$& $1.84\pm0.25$\\
             &11                &                 & $5.10\pm0.33$                      &$5.5\pm0.7$                           &$0.34\pm0.07$        & $1.8\pm0.5$& $1.40\pm0.12$\\  \hline \hline
 
$~^{35}$Cl$^{10+}$   &70                   & 27.5             &         $-$             &   $17.5\pm2.1$                     &$0.97\pm0.09$         & $5.1\pm0.5$  &$57\pm63$     \\ 
	     &77                    &                  &             $-$                        &  $17.1\pm2.4$                & $0.88\pm0.08$      & $4.6\pm0.4$  & $8\pm5$  \\ \hline 
\end{tabular}}
\caption{Data for the different ions incident on PET.
Errors on $REL$ are the standard deviation of values computed along the latent track.
In column~6, the mean value of the semi-minor axis ($b$), normalized to the bulk etch length ($V_Bt$)~\cite{Jeong:2017} is given. 
In column 7 and 8 are the etch-rate ratios measured by the methods discussed in the text.
Errors in column 6, 7 and 8 include statistical and systematic uncertainties.}
\label{ions_PET}
\end{table}
\end{tiny}

\par
It should be mentioned here that there exist 
 innovative ways 
 of increasing the sensitivity of $B$ as the so-called \textquotedblleft two-step etching\textquotedblright~process~\cite{KODAIRA201236}.
 However this method is inapplicable here due to a comparatively shorter range of the incoming particles.
 As shown in Fig.~\ref{Bcut} the 
 normalized semi minor-axis $B$ levels out to $\approx1$ at $REL>17$ MeV/mg cm$^{-2}$ 
 (similar to what observed in CR-39 at $REL\sim6$ MeV/mg cm$^{-2}$~\cite{GIACOMELLI199841}). 
 Such reduced sensitivity of the etch-pit diameter (or minor-axis) at large $REL$s~\cite{BALESTRA2007254} along with
 systematic uncertainties on $a$, $b$ and $V_B$ compels one to switch to the depth measurement method for $V_T/V_B$ determination. 

\subsection*{\bf{Calibration of PET nuclear track detector}}

\begin{figure}[h]
\centering
\includegraphics[width=350px,height=300px]{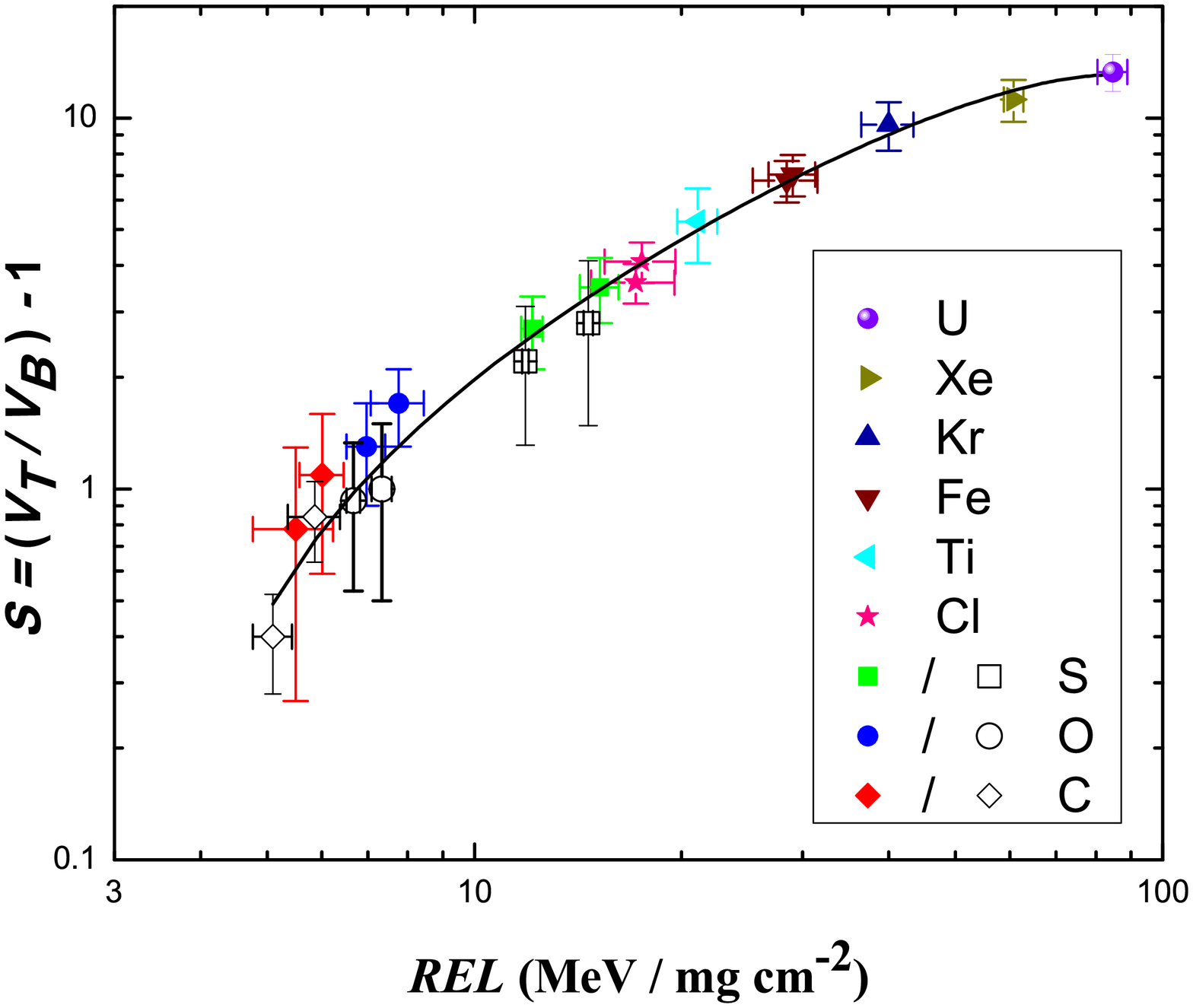}
\caption{Reduced etch-rate data versus $REL$ for PET using etch-pit depth measurements (filled symbols) and diameter measurements (empty symbols).
The curve is the result of the fit to a second order polynomial. The adjusted $R^2$ value is $0.987$.}
\label{PET_calibration} 
\end{figure}

In previous calibration campaigns, PET films had been irradiated with $^{238}$U, 
$^{129}$Xe, $^{78}$Kr, $^{49}$Ti beams at REX-ISOLDE CERN~\cite{DEY2011805}; $^{56}$Fe, $^{32}$S, $^{16}$O beams at 
IUAC~\cite{Dey:2014tza}; $^{12}$C beam at IOP~\cite{BHATTACHARYYA201663}. In Fig.~\ref{PET_calibration}, the reduced etch-rate 
$S=V-1$, is plotted against $REL$ for data listed in Table~\ref{ions_PET} and data obtained from previous exposures. 
On the horizontal axis  $REL$ values are the average over the 
length $L_d$ or $L_h$ for $V_T/V_B$ determined from the depth measurement and the diameter measurement, respectively.
Data are fit to a second-degree 
polynomial equation 
$S=p_0+p_1~(REL)+p_2~{(REL)}^2$, where $p_0=-(1.15\pm0.09)$, $p_1=0.331\pm0.014$ [MeV/mg cm$^{-2}$]$^{-1}$ and  
$p_2=-(19.2\pm2.1)\times10^{-4}$ [MeV/mg cm$^{-2}$]$^{-2}$.  
The fit adjusted $R^2$ is $0.987$. The detection threshold is at $REL\approx3.5$ MeV/mg cm$^{-2}$ determined by extrapolating the curve  
to $S=0$. Therefore PET is able to record particles with $Z/\beta>125$.

\section{Conclusion and discussion}

There is one clear advantage of determining $V_T/V_B$ from the diameter measurement method (Eq.~(\ref{MoEDAL}))
over the depth measurement method (Eq.~(\ref{lab})). 
In the first case, it is relatively easier to focus the microscope on the opening of the etch-pit only, thus reducing 
effectively the time for large area scanning, compared to depth measurement. 
However, as we have shown, the diameter measurement method becomes less sensitive for PET NTD detector when $REL\gtrsim15$ MeV/mg cm$^{-2}$.
Depth measurement provides higher resolution at larger $REL$ values.
In conclusion in identifying a particle using nuclear track detectors, depth measurement method for determining $V_T/V_B$ should be 
adopted if normalized semi minor-axis $B\approx1$.

 \section*{Acknowledgements}
 
 We sincerely thank the staffs at the Inter-University Accelerator Center (IUAC),
New Delhi, India, especially Mr. N. Saneesh and Mr. Mohit Kumarfor, for providing all 
possible support during the chlorine beam exposure. The authors also thank Mr. Sujit K. Basu for technical assistance.
The work is funded by IRHPA (Intensification of Research in
High Priority Areas) Project (IR/S2/PF-01/2011 dated 26.06.2012)
of the Science and Engineering Research Council (SERC), DST,
Government of India, New Delhi. LP and VT wish to thank their colleagues at INFN Bologna.

\bibliography{mybibfile}

\end{document}